\documentclass[twocolumn,prl,floats,superscriptaddress]{revtex4-2}

\usepackage[T1]{fontenc}
\usepackage[utf8]{inputenc}
\usepackage{graphicx}
\usepackage{amsmath}
\usepackage{amssymb}
\usepackage{color}
\usepackage{xspace}
\usepackage{multirow}
\usepackage{hyperref}
\usepackage{dcolumn}
\begin{document}
\title{Evidence for Temperature Chaos in Spin Glasses}

\author{Qiang Zhai}
\email{qiang@utexas.edu}
\affiliation{Texas Materials Institute, The University of Texas at Austin, Austin, 
Texas  78712, USA}
\author{Raymond~L.~Orbach}
\email{orbach@utexas.edu}
\affiliation{Texas Materials Institute, The University of Texas at Austin,
  Austin, Texas  78712, USA}
\author{Deborah L. Schlagel}
\affiliation{Division of Materials Science and Engineering, Ames Laboratory,
  Ames, Iowa, 50011, USA}

\date{\today}
\begin{abstract}
 We study the field cooled magnetization of a CuMn spin glass under temperature perturbations. The $T$-cycling curves are compared with the reference curve without temperature cycling. There is a crossover from  the cumulative aging region to noncumulative aging  region as the temperature change is increased. The cumulative aging range scales with the chaos length $\ell_\mathrm{c}$, becoming comparable to the correlation length, $\xi$,  at the crossover boundary.  The extracted chaos exponent, $\zeta = 1.1$,  is in agreement with theoretical predictions. Our results strongly suggest temperature chaos exists in real spin glasses system.
\end{abstract}
\maketitle
The equilibrium spin configuration in the spin glass phase \cite{mezard:1987,fisher:1986a,fisher:1988} is predicted to reorient for arbitrarily small temperature variations on a length scale greater than, $\ell_\mathrm{c}(T_1, T_2)$, the chaos length \cite{fisher:1988, bray:1987}.  The temperature chaos (TC) effect was first exhibited through a renormalization group approach \cite{mckay:1982,banavar:1987} and $T=0$ fixed point scaling arguments \cite{bray:1987, fisher:1988}. It has been studied analytically \cite{nifle:1992,kondor:1993,neynifle:1993,cugliandolo:1999, rizzo:2006,barucca:2014,panchenko:2016,chen:2017,arous:2020,eldan:2020} and observed through simulations \cite{rieger:1996,neynifle:1997,neynifle:1998,almeida:2013} on different lattice structures \cite{cieplak:1993,huse:1997, sasaki:2000,yoshino:2001,krzakala:2004,lukic:2006,parisi:2010,thomas:2011,sun:2012} by utilizing  improved computable lattice sizes \cite{billoire:2002} and strategies \cite{krzakala:2002, katzgraber:2007,jorg:2012,fernandez:2013,billoire:2014,wang:2015,billoire:2018}, in both droplet-like \cite{yoshino:2003,sasaki:2003,takayama:2004,sasaki:2005} and replica symmetry breaking scenarios \cite{sasaki:2002b, rizzo:2003}. Apart from the many results obtained at equilibrium, a recent simulation study \cite{baity:2021}  exhibited the evidence of TC in nonequilibrium states using the Edwards-Anderson Ising  spin glass model. \\

In contrast with the relevant accomplishments in theoretical studies, the experimental verification  \cite{sandlund:1988, mattsson:1993, jonason:1998, jonsson:1999,jonsson:2002,arai:2007} of TC remains controversial \cite{alers:1992,komori:2000, bouchaud:2001, berthier:2003}. Because of sluggish glassy dynamics,  experiments on spin glasses work exclusively in a nonequilibrium regime, accompanied by aging (logarithmically slow relaxation) \cite{lefloch:1992,vincent:1995,picco:2001b,suzuki:2004}, rejuvenation \cite{vincent:2000,vincent:2009}, and memory effects (recovery of the response generated at a temperature stop before successively cooling and  heating back to the original temperature) \cite{jonason:2000,mathieu:2001,sasaki:2002}.  It has been argued that rejuvenation, a renewal of the aging process (or a discontinuity in exploring phase space), can occur without invoking TC. Doubts \cite{scheffler:2003,jonsson:2003,jonsson:2004} have been cast on these latter claims \cite{berthier:2002a,berthier:2002b,sales:2003,berthier:2005}, because such findings could be obtained by tuning of parameters \cite{krzakala:2006} or be camouflaged by small timescales, lattice size \cite{aspelmeier:2002},  and cooling rates effects \cite{picco:2001} in simulations.  It has been argued further  that a second rejuvenation \cite{takayama:2002,krzakala:2006}, under the protocol when the system is cooled to a lower temperature and then heated back, should be understood within the formalism of TC. A well-established picture of TC in spin glasses will contribute to a better understanding of similar phenomenon in more general random systems \cite{mcnamara:1999, sales:2002,da:2004,le:2006,kustov:2019,da:2021} and developing quantum annealers \cite{katzgraber:2015,victor:2015, vinci:2015,zhu:2016}. \\

 Here, we take rejuvenation  \cite{hammann:1992, jonsson:2002} in the aging process of the field cooled (FC) magnetization as a sign of the onset of TC. The experimental  protocol we employ involves a  temperature cycling process, whence the rejuvenation effect should be classified as second rejuvenation.  We calibrate the chaos length using the correlation length, $\xi$, a measure of the size of glassy domains \cite{rieger:1994}, controlled by the varied initial temperature $T_1$ and fixed aging time $t_1$. The scaling relation between the chaos length and the \emph{reversible temperature range} $\delta T^\text{rev}$, where the aging is cumulative between two temperatures in a $T$-cycling process, is shown to be satisfied. A chaos exponent $\zeta$ \cite{bray:1987} close to the theoretical value of unity is extracted. These results strongly suggests the existence of TC in the canonical CuMn spin glass. \\

\emph {Experimental design. }   The underlying premise behind our experiments is based on a length scale argument. In the absence of TC,  aging at  two temperatures $T_1$ and $T_2$ contributes cumulatively to the growth of the correlation length at each temperature. In the presence of TC, the correlation growth at $T_1$ cannot be projected onto the growth of the correlation length at $T_2$, and vice versa.  For example, consider aging at $T_1$ to generate a certain  correlation length, $\xi(T_1)$. Upon changing the temperature to $T_2$,  instantaneously $\xi(T_2) = \xi(T_1)$.  If $\ell_\mathrm{c}(T_1, T_2) /\xi(T_1) \geq 1$, aging at $T_2$ continues the correlation length growth from $\xi(T_1)$.  If, however,  $\ell_\mathrm{c}(T_1, T_2) / \xi(T_1) < 1$, the system experiences incoherent spin flipping dynamics.  In $T$-cycling experiments, the situation is further complicated because, during the process of heating the system back to $T_1$,  $\xi(t)$ is changing with time, while the chaos length $\ell_\mathrm{c}(T_1, T_2)$ is fixed by the temperature separation. This complication will be discussed in detail below.  \\

We chose working with the FC magnetization \cite{djurberg:1999}, the magnetization response of a sample aged in a constant field without field change after being cooled to $T_1$ from above $T_\mathrm{g}$,  rather than conventional alternatives like zero field cooled (ZFC) \cite{granberg:1988,granberg:1990} magnetization or thermoremnant  (TRM) magnetization \cite{hammann:1992, mathieu:2010}. This is to eliminate the possible chaos induced by a magnetic field change \cite{kondor:1989,billoire:2003}.\\

 Our specific protocol consists of two parts, one to measure the field cooled magnetization, $M_\text{FC}(t,T_1,H)$, without a temperature perturbation as the \emph{reference curve}; and the other, temperature cycling to study cumulative aging and TC.   The reference curve was measured by cooling the sample from 40 K, well above $T_\mathrm{g}$, the spin glass condensation temperature, to the measurement temperature, $T_1$, in a constant field $H=40$ Oe, at 10 K/min. After the temperature was stabilized at $T_1$ for 100 s, the reference curve magnetization, $M_\text{FC}^\text{ref}(t,T_1)$, was recorded. \\

In the $T$-cycling experiment, the sample was cooled from 40 K to $T_1$ at 10 K/min, and the magnetization recorded under the same temperature stabilization condition for a duration of  $10^4$ s. For clarity, the first aging period is denoted as $t_1$, and the instantaneous point in time before dropping the temperature further is referred to as $\hat{t}_1$. After $\hat{t}_1$, the temperature was lowered to $T_2=T_1 - \Delta T$. The temperature cooling rate was adjusted to lie between a range of $8$ mK/min to $100$ mK/min in order to minimize temperature down-shoot after reaching $T_2$.  After $t_2^{(1)}=10^3$ s aging at $T_2$, the sample was heated back with a symmetrical heating profile to $T_1$, again avoiding temperature overshoot.\\
 
It should be noted that because of the finite cooling rates, the correlation length will continue to grow during the cooling and heating process. The time spent in the cooling and heating process between $T_1$ and $T_2$ is denoted as $t_2^{(2)}$, of the order of 100 s. Together with $t_2^{(1)}$, the total time spent during temperature cycling would be $t_2 = t_2^{(1)}+t_2^{(2)}$. The magnetization was recorded 100 s after the temperature was stabilized at $T_1$,  during the period $t_3$, and was denoted as $M^\text{cyl}_\text{FC}(T_1, t_3; T_2, t_2; T_1, t_1)$. It should be emphasized that the magnetization measurements were conducted during $t_1$ and $t_3$  at $T_1$ only. The time and temperature profile as a function of time are plotted in Fig. 1 in the Supplemental Materials (SM).\\

One needs an appropriate method to characterize the reference curve,  and the $T$-cycling curve, in order to interpret the experimental results.  Traditionally,  a collapse of a family of curves to a master curve \cite{hammann:1992}, and the effective aging time \cite{jonsson:2002}, $t_\mathrm{w}^\mathrm{eff}$ (defined as the time when the relaxation curve, $S(t)=\mathrm{d}M(t)/\mathrm{d}\mathrm{log}(t)$, peaks),   have been used to characterize the aging curves. The former needs additional parameterization in order to convert the physical time into reduced time \cite{alba:1986}. For the latter, the time at which $S(t)$ peaks becomes more difficult to extract for long waiting times,  as the $S(t)$ curve broadens substantially.  Moreover, the relaxation curve is measured under a change of magnetic field.  The effective aging time is reduced as a function of field and correlation length \cite{bert:2004, zhai:2017}, leaving the assumption that $t_\mathrm{w} \sim t_\mathrm{w}^\mathrm{eff}$ \cite{jonsson:2002} to be somewhat precarious for large values of $\xi$, where $t_\mathrm{w}$ is the time before a field is switched on or off. This may explain why  the data points do not overlap well  at the timescale of $10^5$ s in the cumulative aging range reported in Ref. \cite{jonsson:2002}.  Along with possible magnetic field induced chaos,  these factors may be responsible for their small value of $\zeta$  that lies well outside the theoretical estimates.  \\

 Because TC has been reported to be a subtle \cite{rizzo:2003} and gradual \cite{fernandez:2013} effect, we have chosen to overlap the cycling curve and reference curves directly. The drawback of our approach is the vulnerability of signal quality to external noise. A squid signal jump caused by environment noise ruins a set of measurements. Note that before $\hat{t}_1$, there is no difference in protocols between the reference curve and the $T$-cycling curve measurements, $M^\text{cyl}_\text{FC}(T_1, t_1)$ and $M^\text{ref}_\text{FC}(t)$ can be overlapped directly. After $\hat{t}_1$, if a time translation $\delta t$ can be found, such that $M^\text{cyl}_\text{FC}(T_1, t_3+\delta t; T_2, t_2;T_1, t_1) = M_\text{FC}^\text{ref}(t,T_1)$, the aging can be taken as cumulative and reversible. Otherwise, when the overlap process fails,  rejuvenation is signaled as a consequence of TC.  \\

The sample used in our experiments is a 6 at.\ \% CuMn single crystal \cite{zhai:2019} grown at the Ames Lab with a spin glass transition temperature $T_\mathrm{g}=31.5$ K. Details of the preparation of the sample can be found in the SM.  The  magnetization measurements were taken on a Quantum Design commercial superconducting quantum interference device. The temperature sensor is located within 5 mm of the sample.\\

\begin{figure*}[t]
\centering
\includegraphics[width=1\columnwidth]{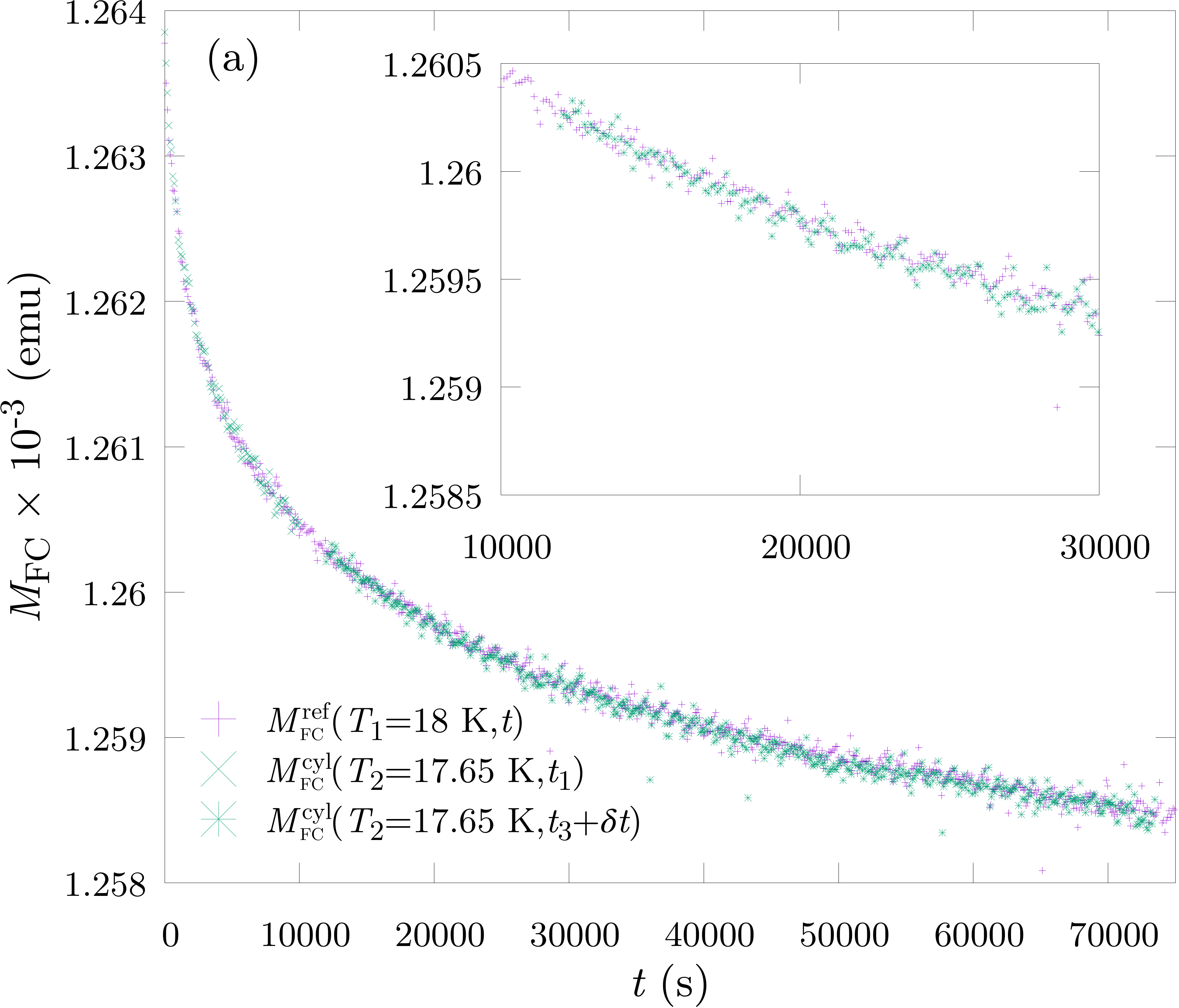}
\includegraphics[width=1\columnwidth]{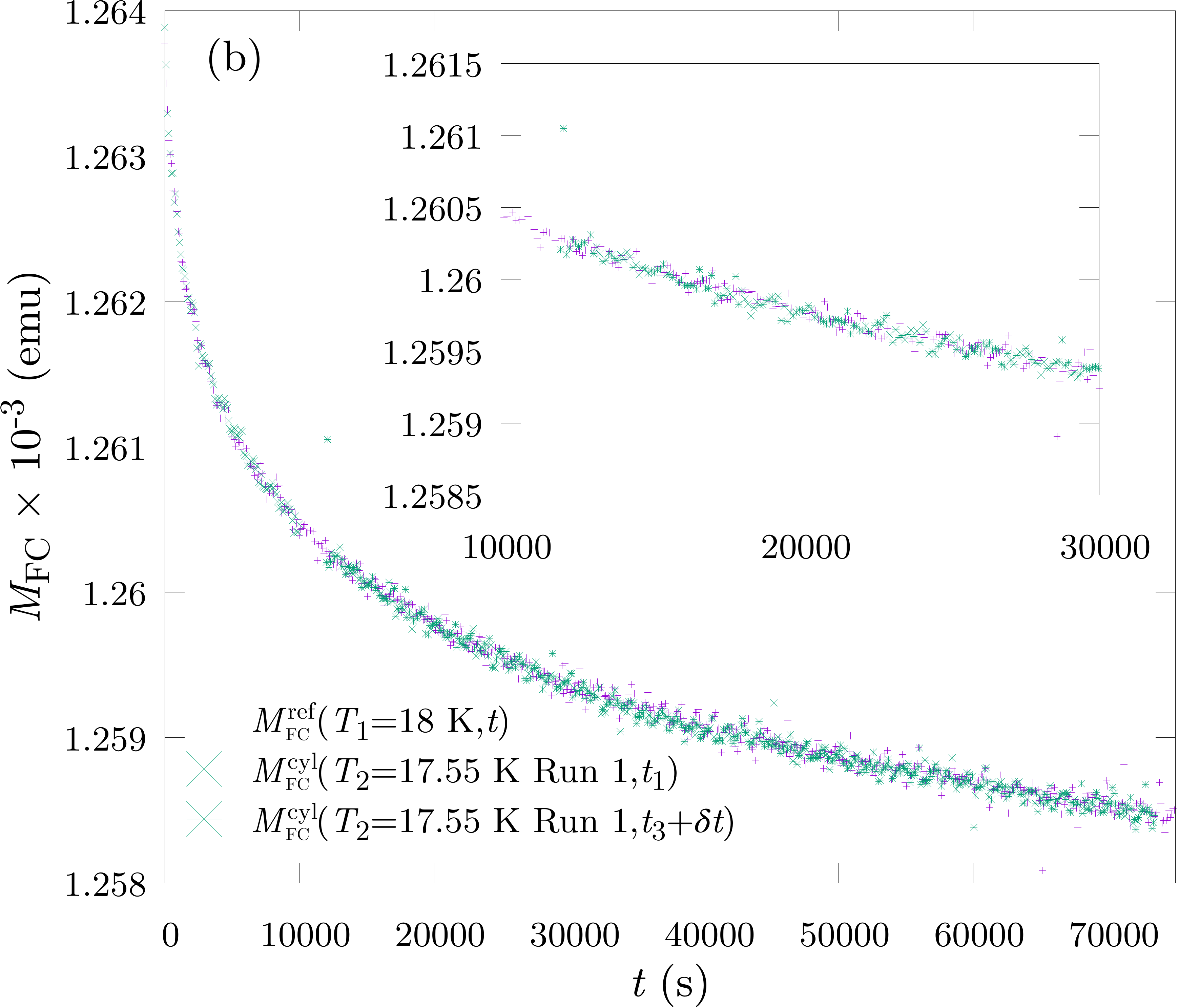}
\includegraphics[width=1\columnwidth]{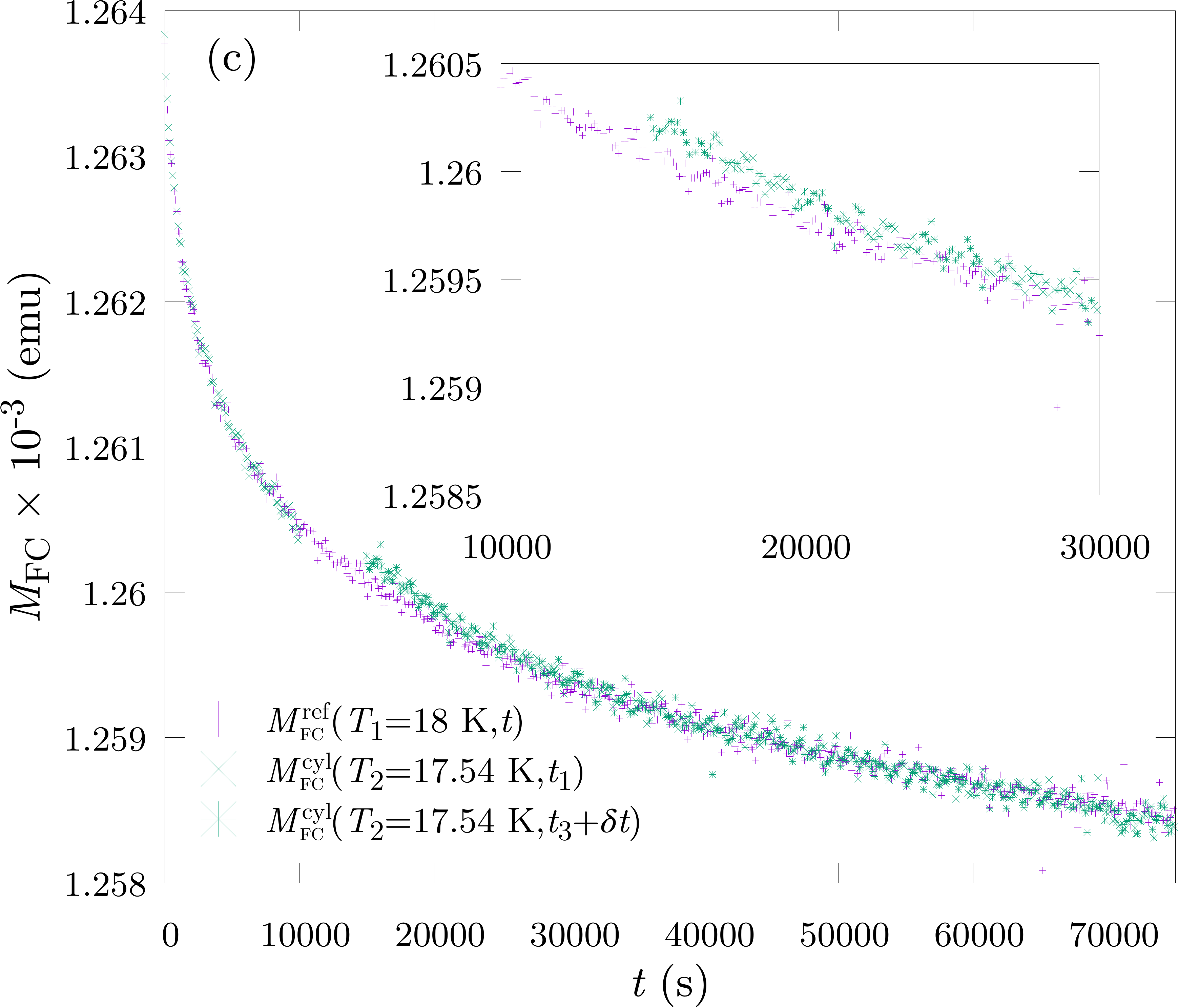}
\includegraphics[width=1\columnwidth]{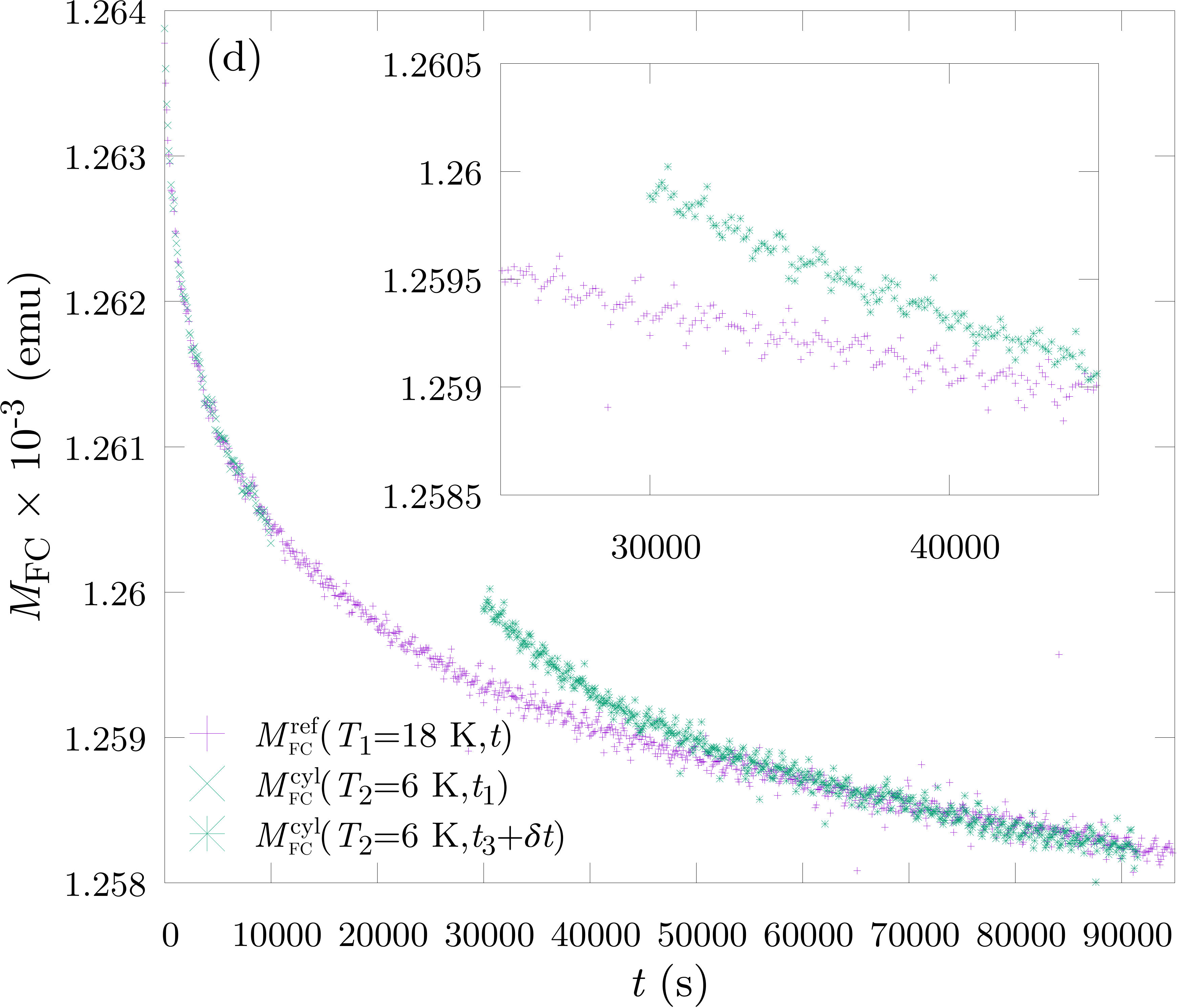}
\caption{The example of $T_1=18$ K. The temperature is gradually lowered to $T_2$ after $t_1\approx10^4$ s, and heated back after $t_2^{(1)} \approx 10^3$ s. The $T$-cycling curve is then shifted by $\delta t$ to overlap the reference curve. In the reversible temperature range, (a) and (b), the cycling curve can be overlapped with the reference curve over the whole period $t_3 \approx 7\times 10^4$ s. In the chaotic range, (c) and (d), the cycling curve can only be partially overlapped. Thus, we conclude that, at $T=18$ K, TC sets in for $\Delta T > 450$ mK.}
\label{fig:18K}
\end{figure*}

\emph{The transition to temperature chaos.} In order to observe the transition from cumulative aging to noncumulative aging (i.e., temperature chaos), we fixed $T_1$, and gradually lowered the temperature $T_2 = T_1 -\Delta T$, in a series of measurements, until the $T$-cycling curves could no longer be overlapped with the reference curve.  Initially, the $\Delta T$ was reduced at 100 mK steps to locate a rough onset of TC. Because the temperature resolution of our measurement was $\pm$ 5 mK,   a 10 mK temperature step was used thereafter to determine a more accurate boundary between reversible cumulative aging  and noncumulative aging. A typical example is shown in Fig. \ref{fig:18K}  at  18 K, where the approximate magnitude of reversible range is $\delta T^\text{rev} = 450$ mK.  The same process was repeated for different values of $T_1$, down to 14 K.  The  reversible cumulative aging temperature ranges are listed in Table \ref{tab:trev}.   A  complete set of data for all the measured temperatures can be found in the SM. \\

\begin{table}
\begin{ruledtabular}
\begin{tabular}{c  c  c  c  c   c}
$T_1 $ (K) &18&17 & 16&15 &14 \\
$\delta T^\text{rev}$ (mK)&450 &510 &600 &690 &800\\
$t_2^{(T_1)}$ (s)&2000&2000&2500&3000&1500\\
\end{tabular}
\end{ruledtabular}
\caption{The extracted reversible cumulative aging range $\delta T^\text{rev}$ and the effective aging time $t_2^{(T_1)}$ for different $T_1$. The data are sampled every 100 s, the time translation used to overlap to the reference curve is multiples of 100 s. }
\label{tab:trev}
\end{table}

The underlying physical process can be described by a competition between the correlation length $\xi$, and the chaos length $\ell_\mathrm{c}(T_1, T_2 = T_1-\Delta T)$. After aging at $T_1$ for a period of $t_1$, the correlation length reaches $\xi(T_1,t_1)$, as described by either power law growth \cite{rieger:1994,zhai:2017b}, or logarithmic growth \cite{jonsson:2002b} for the droplet-scaling model \cite{fisher:1986a}. For brevity, we consider power law dynamics, with a similar analysis for the other. Following a temperature drop, if  
\begin{equation}\label{x1}
x_1 = \ell_\mathrm{c}(T_1, T_2)/\xi(T_1,t_1)\ge1~,
\end{equation}
the correlation length will continue to grow. The time taken in the cooling process has been omitted in the above expression. \\

However, one should be aware that the aging rates of $\xi$ at $T_1$ and $T_2$ are, in general, different. According to the power law growth rate, $\xi(t, T) = b(t/\tau_0)^{1/z(T)}$, where $b$ is a geometrical factor,  $1/\tau_0 \sim k_\mathrm{B}T_\mathrm{g}/\hbar$ is the exchange attempt frequency, and the exponent $z(T)$ sets the aging rate. The factor, $z(T)$ \cite{baity:2018},  is linearly dependent upon $T$, and has the form $z(T) = z_\mathrm{c}T_\mathrm{g}/T$, where $z_\mathrm{c}$ \cite{zhai:2019} can be approximately taken as a constant over our temperature range. For cumulative aging, Eq. \eqref{x1}, the aging time $t_1$ at $T_1$ can be effectively converted to a aging time $t_1^*$ at $T_2$ through
\begin{equation}
(t_1/\tau_0)^{T_1} = (t_1^{*}/ \tau_0)^{T_2}.
\end{equation}
After the aging period $t_2$ at $T_2$, the correlation length reaches approximately,
\begin{equation}
\xi(T_2, t_2;T_1, t_1) \approx b[(t_1^*+t_2^*)/\tau_0]^{T_2/(z_\mathrm{c}T_\mathrm{g})}~,
\end{equation}
in which $t_2^{(2)}$ can be projected to an effective aging time $t_2^{(2*)}$ at $T_2$, and $t_2^* = t_2^{(1)}+t_2^{(2*)}$, if  Eq.\eqref{x1}, holds. \\

Whether aging is cumulative or not, upon heating the system back to $T_1$, depends on the variable,
\begin{equation}\label{x2}
x_2 = \ell_c(T_1, T_2)/\xi(T_2, t_2;T_1, t_1)~.
\end{equation}
Again, if $x_2\ge1$, the  correlation length growth continues, and cumulative aging results in an overlap of the reference curve and the cycling curve over the full experimental time period. Either the opposite of Eq.\eqref{x1}, or $x_2 <1$ in Eq.\eqref{x2}, will result in a chaotic reorientation of the spin configuration. Thus, in general, the measured reversible range $\delta T^\text{rev}$ is a function of ($T_1$, $t_1$) that sets the correlation length before the temperature drop to $T_2$,  and ($T_2$, $t_2$) that sets the correlation length growth during the temperature cycling period. \\

\emph{Extraction of the chaos  exponent $\zeta$.} TC was initially introduced through a renormalization group argument. In that context, we shall use the associated scaling arguments for the extraction of $\zeta$. The chaos length (overlap length) $\ell_\mathrm{c}(T_1, T_2)$ is the length scale within which the spin correlations are free of the influence of temperature variation. For small temperature changes, the free energy at temperature $T_1$ \cite{fisher:1988} is approximated from the free energy at $T_2$ through a Taylor expansion,
\begin{equation}\label{eq:ap1}
F(T_1) \approx F(T_2)-(T_1-T_2)S(T_2),
\end {equation}
where $F(T) = \gamma(T)\ell^\theta$, and $S(T) = \sigma(T)\ell^{d_\mathrm{s}/2}$, the droplet interface free energy and entropy,  respectively. Here, $d_\mathrm{s}$ is the surface fractal dimension \cite{fisher:1988}.  Alternatively, noticing that the energy term is similar at the two temperatures at the chaos length scale \cite{katzgraber:2007},
 \begin{equation}\label{eq:ap2}
F(T_1) \approx F(T_2)+T_2S(T_2)-T_1S(T_1).
\end {equation}
Upon a sign change of the free energy at $T_1$, the above approximations in Eqs. (\ref{eq:ap1} , \ref{eq:ap2}) yields
\begin{equation}\label{eq:lc}
\begin{aligned}
\ell_\mathrm{c}(T_1, T_2) &= \left[\frac{\gamma(T_2)}{(T_1-T_2)\sigma(T_2)}\right]^{1/(d_\mathrm{s}/2-\theta)},\\
\ell_\mathrm{c}(T_1, T_2) &= \left[\frac{\gamma(T_2)}{T_1\sigma(T_1)-T_2\sigma(T_2)}\right]^{1/(d_\mathrm{s}/2-\theta)}.
\end{aligned}
\end{equation}
Thus, the chaos exponent, $\zeta =d_\mathrm{s}/2-\theta$, and $S(T) $ is proportional to $\sqrt{T}$ \cite{aspelmeier:2002}.\\

\begin{figure}
\centering
\par\smallskip
\includegraphics[width=\columnwidth]{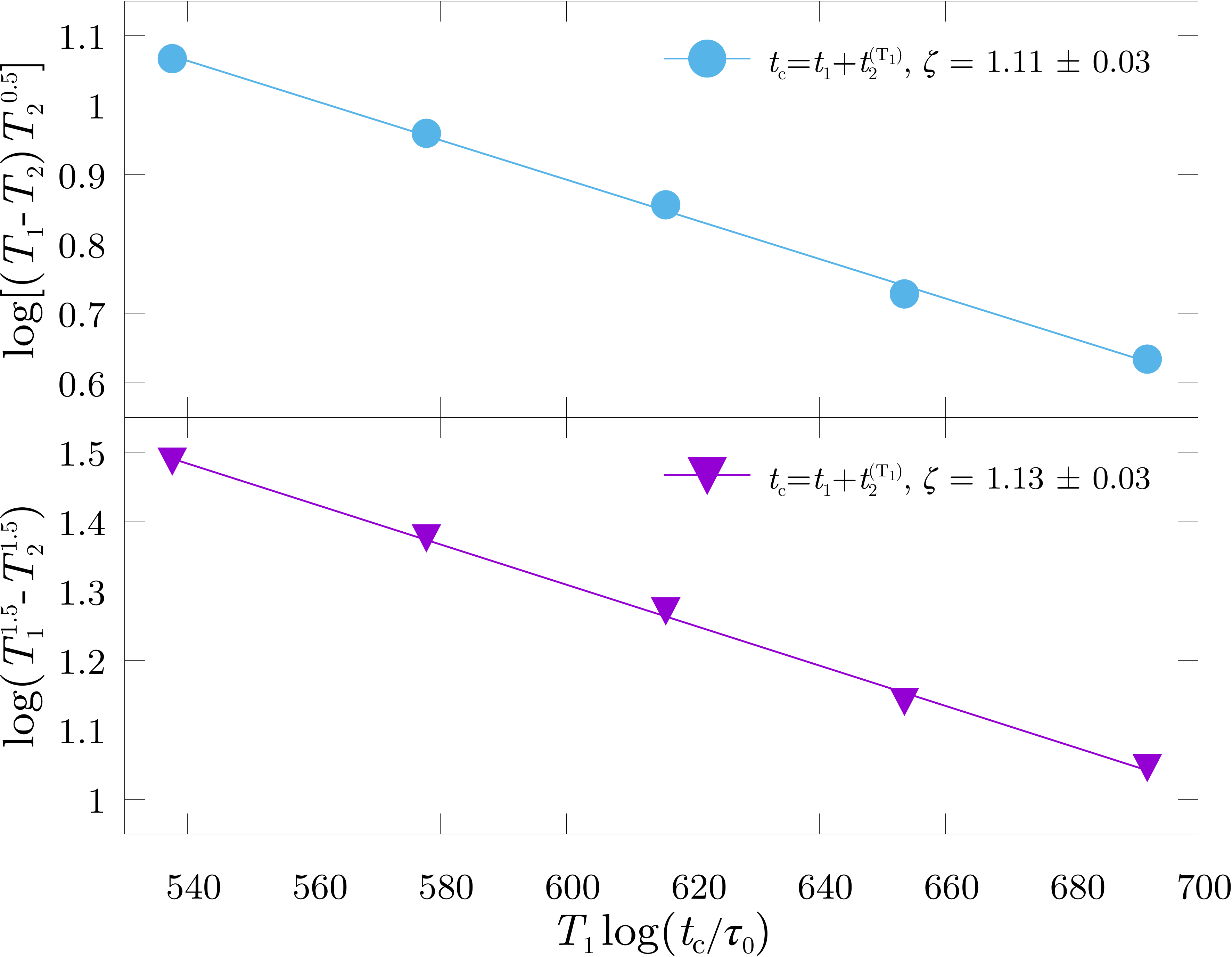}
\caption{Using a power law growth for $\xi$, the chaos exponent $\zeta \approx 1.1$ is extracted for different approximation schemes following Eq. \eqref{eq:lc}.}
\label{fig:power}
\end{figure}

To extract a value for $\zeta$, we assume that the length scale $\ell_\mathrm{c}(T_1, T_2) \approx \xi_\mathrm{c}$ at the boundary between cumulative aging and noncumulative aging. Here, $\xi_\mathrm{c}$ is the correlation length when the system is heated back to $T_1$,  and $T_1 - T_2 = \delta T^\text{rev}(T_1, T_2)$.  For power law growth, $\xi_\mathrm{c} = ba_0(t_\mathrm{c}/\tau_0)^{[T/(z_\mathrm{c}T_\mathrm{g})]}$, where $t_\mathrm{c}$ is the timescale used to estimate $\xi_\mathrm{c}$. Recall that, in the reversible range, the cycling curve can be overlapped with the reference curve after a time translation.  The time difference between the first point of the translated curve measured during $t_3$ and $\hat{t}_1$ is denoted by, $t_2^{(T_1)}$, the time taken as the effective aging time over the period of $t_2$. The sum of $t_2^{(T_1)}$  and $t_1$ is used for $t_\mathrm{c}$ at the crossover boundary (the lowest temperature drop without rejuvenation).  The exponent $z_\mathrm{c} =12.37$ \cite{zhai:2019} was measured on the same sample. The scaling relation between $\ell_\mathrm{c}( \xi_\mathrm{c})$ and the temperature drop, following Eq.\eqref{eq:lc}, is exhibited in Fig. \ref{fig:power}, with $\zeta \approx 1.1$ for two different approximation schemes. This value is within the range of  well accepted values for $\zeta$ that scatter around unity from theoretical calculations and simulations \cite{feng:1987,katzgraber:2007,fernandez:2013,wang:2015,baity:2021} for the 3D Ising model. Specific values are provided in Table I of the SM.\\

\begin{figure}
\centering
\includegraphics[width=\columnwidth]{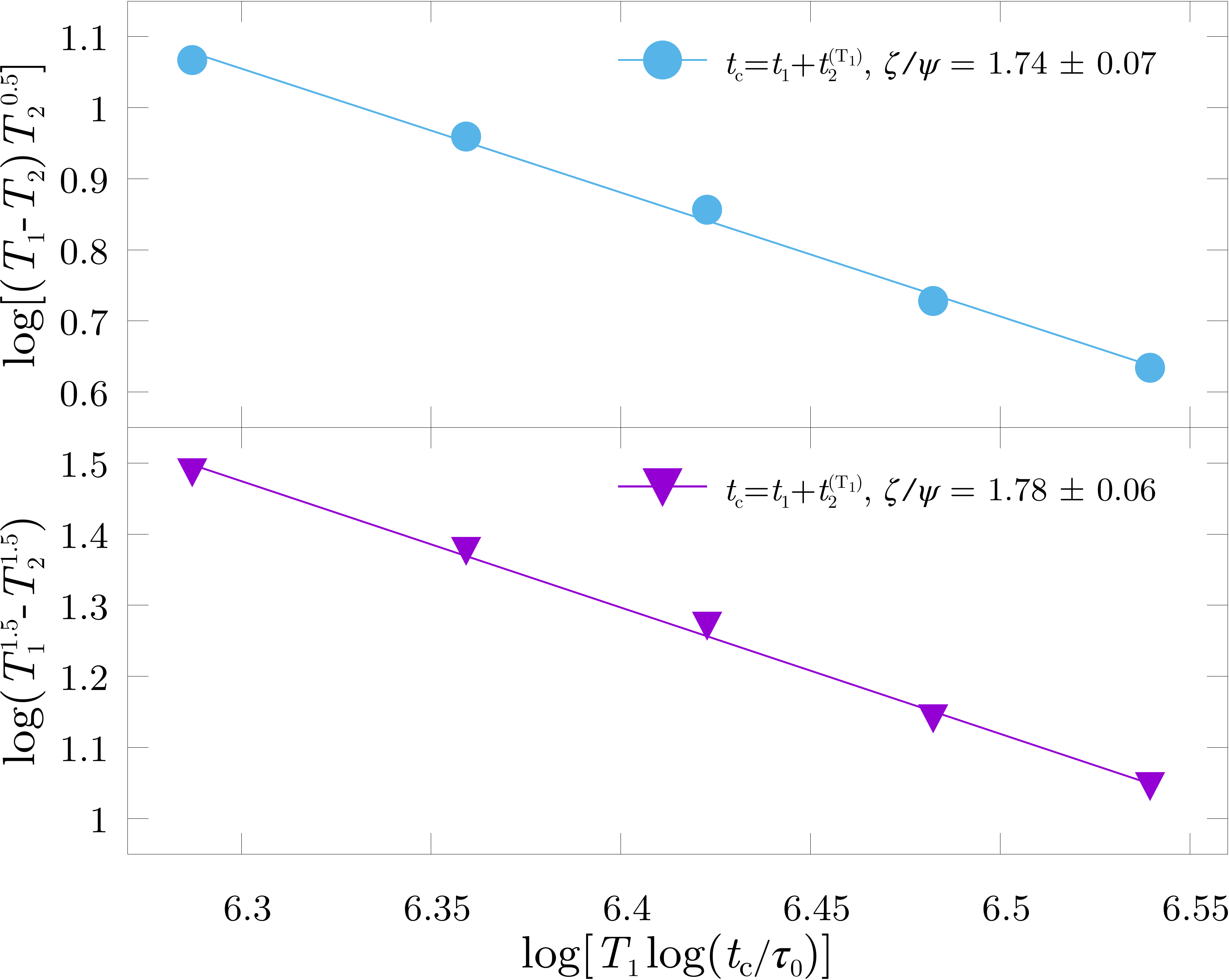}
\caption{Using a logarithmic growth law of $\xi$, the ratio of $\zeta/\psi \approx 1.7$ compared with the power law growth result shown in Fig. \ref{fig:power}.}
\label{fig:log}
\end{figure}

For logarithmic growth, the correlation length would grow as, 
$\xi_\mathrm{c} = ba_0[\mathrm{log}(t_\mathrm{c}/\tau_0)k_BT/\Delta(T)]^{1/\psi}$, where $\Delta(T)$ is the free energy barrier for the thermal activated process \cite{jonsson:2002}. Using the same choice of $t_\mathrm{c}$, and assuming $\Delta(T)$ does not vary too much,  the fitted curve in Fig. \ref{fig:log} gives the ratio $\zeta/\psi \approx 1.7$. By requiring $\zeta = 1.1$, $\psi \approx 0.65$, in the range of reported values \cite{dupuis:2001,bert:2004} for  the droplet model. \\

\emph{Discussion and Conclusion}. The concept of TC was introduced in the context of equilibrium properties \cite{bray:1987,fisher:1988}. Nevertheless, the glassy domain characterized by the correlation length formed towards equilibrium  should be indistinguishable from that in equilibrium. Moreover, the concept of TC has been shown to exist under nonequilibrium condition \cite{baity:2021}. Further, the idea of TC is compatible with other prominent theories of spin glasses \cite{sasaki:2002b, fernandez:2013,baity:2021}. Beyond that, TC is widely reported in more general  frustrated complex systems, for example, with $\zeta=1$ for the disordered Bose fluid \cite{da:2021}. Although the value of $\zeta$ appears to vary for different physical systems (see a different reported value for elastic media \cite{mcnamara:1999}). The scaling law between,  $\ell_\mathrm{c}$, the chaos length  and  an external perturbation may well be general. \\

Though CuMn is often categorized as a metallic Heisenberg system,  the introduction of anisotropy (because, e.g., of the omnipresent Dzyaloshinskii-Moriya interaction) \cite{martinmayor:2011,baity:2014} breaks the rotational symmetry and makes the system one of the Ising universality.  In addition, Ising-behavior of CuMn has been  affirmed through the synergy between experiments and simulations \cite{zhai:2020, paga:2021}.\\

With only one free variable, $z_\mathrm{c}$, in the power-law approximation (actually a measured value for the identical sample), the closeness between the extracted chaos exponent and theoretical values from the Ising model calculations and simulations is striking. The crossover from cumulative aging  to noncumulative aging  is therefore well explained within the framework of TC.  We believe our findings provide strong evidences for TC in a real spin glass system. \\

\begin{acknowledgments}
We thank helpful communications with Victor Martin-Mayor. This work was supported by the U.S. Department of Energy, Office of Basic Energy Science, under Award No. DE-SC0013599.\\
\end{acknowledgments}

%

\end{document}